\documentclass[prd,preprint,11pt,nofootinbib]{revtex4}
\usepackage{epsfig}

\begin{document}
\def\D0{\mbox{D\O}}

\renewcommand{\thefootnote}{\fnsymbol{footnote}}

\title{\vspace{3in}
\LARGE{Direct search limits on the Littlest Higgs model}}
\author{\Large{John Boersma\footnotemark[1]}}
\affiliation{\Large{Department of Physics and Astronomy,\\
University of Rochester\\
Rochester, NY 14627-0171\\}\vspace{2.cm}}

\begin{abstract}
Recent direct searches for new massive particles place constraints on the free parameters
of the Littlest Higgs model.  Depending on the choice of model free parameters,
the direct search limit on the global symmetry breaking scale $f$ can range from as low as a few 
hundred GeV to in excess of $4.5$ TeV. The most stringent constraints are from exclusion of the $A_H$
using high-mass dilepton resonance searches.  The $Z_H$ provides the best constraint in parameter
regions where the $A_H$ decouples from leptons.  Current top pair resonance data approach but do not yet 
reach a useful limit in the anomaly-cancelling case, but do provide a constraint for a limited
range of parameters in other cases.   
A neutral gauge boson is shown to be undetectable
in dilepton resonances for a significant range of parameter space due to decoupling from Standard Model leptons,  
providing a counterexample to broad claims
that a new neutral gauge boson (sometimes generically referred to as a $Z'$) is ruled out to a high mass scale.

\end{abstract}

\footnotetext[1]{jboersma@pas.rochester.edu}

\maketitle

\renewcommand{\thefootnote}{\arabic{footnote}}

\section{INTRODUCTION}

The Standard Model of particle physics continues to meet all experimental
tests, and yet studies of the structure
of the model suggest that it is incomplete.  One well-known difficulty is that the mass of the
Higgs boson, an essential participant in the standard description of electroweak interactions,
receives loop corrections to its mass (squared) that are quadratic in the
loop momenta. The largest correction is due to the top quark loop, with smaller contributions
coming from loops of the electroweak gauge bosons and of the Higgs boson itself. 
These corrections offset a tree-level mass term for the Higgs boson. 
At a high energy scale, the loop contributions become large, and 
require a precise near-cancellation
by the tree-level mass term in order to achieve an effective
Higgs boson mass on the order of $100$ GeV, as required by fits to precision electroweak parameters~\cite{lep}.
The existence of such a near-cancellation is 
highly suggestive of an as-yet unidentified partial or broken symmetry.  This symmetry, evident
at high energy scales, would be the cause of a cancellation which, at low energy scales, appears only as
an apparent coincidence. 

In supersymmetric models, the problem of quadratic Higgs mass divergences is 
resolved by the introduction of an opposite-statistics partner 
for each particle in the Standard Model.   
Recently, extended symmetry groups have been found which can contain the $SU(2)_L\otimes U(1)_Y$ electroweak
gauge group of the Standard Model as well as additional structure that provides for quadratic divergence
cancellation between {\slshape{same-}}statistics particles.  These are known as ``Little Higgs"
models~\cite{reviews}.  The Little Higgs models realize a long-standing conception of the Higgs as 
a pseudo-Goldstone boson~\cite{pseudo}.  

The ``Little Higgs mechanism" embodied in these models provides an alternative to supersymmetry
for resolving the Higgs mass problem.
The Little Higgs models remove the need for finely-tuned cancellations
to beyond about $10$ TeV, largely beyond the reach of current electroweak constraints on new strong interactions. 
However, Little Higgs models typically leave an uncancelled logarithmic
mass contribution, which requires additional new contributions at some high scale
to preserve a small Higgs mass.  Consequently we need not, and in fact cannot, view a Little 
Higgs model as complete, but must keep in mind that
at sufficiently high energies there will be additional contributions from the unidentified complete theory, the
so-called ``ultraviolet completion"~\cite{uv}.   

A minimum requirement for any proposed extension to the Standard Model is that it conform to known
experimental constraints.  In practice this means that it must not violate the current experimental limits
on the properties of known particles, including fits to the free parameters of the electroweak theory, and the
current search limits for new particles.  Applying this requirement to the Little Higgs models is considerably
complicated by the incomplete nature of the theories.  We may find that the identified contributions
of a certain model violate, say, the known experimental limits on the mass of the $Z^0$ gauge boson, and yet 
not be able to determine whether this violation may be offset by contributions from the ultraviolet completion.
Considerable effort has been made to determine the true constraints to Little Higgs models from precision electroweak
measurements, using a variety of approaches~\cite{Hewett,ept,Chen,KR,Csaki}\footnote{In fact not all such studies agree.  
One study\cite{Chen} concludes, for example at $f=3$ TeV, that only a tightly constrained region of parameter space
with $c > 0.9$ is allowed, while \cite{KR} finds that a wide range of parameter space is allowed, all with $c < 
0.85$. 
}.
In this work, a complementary
effort is made to constrain the Littlest Higgs model using direct search results.  

In Section II of this paper, the Littlest Higgs model is briefly described, primarily to introduce notation.  Section III
presents the main results of the paper, including a scan for the lightest new particles in various regions
of parameter space, the full mass-dependent partial decay widths (at tree level) of the most-often lightest particle, the 
$A_H$, and the combined results of various constraints on the symmetry-breaking scale $f$.  Section IV provides
some conclusions.

\section{THE LITTLEST HIGGS MODEL}

The prototype Little Higgs model is the ``Littlest Higgs"~\cite{LH}. This model is presented in detail, with
Feynman rules and some phenomenological results, in~\cite{Han}. Two expressions for masses are corrected in~\cite{Buras}. 
In this section I will sketch the structure of the
model, and introduce the notation that will be used.

\begin{table}[t]
\begin{tabular}{|c|c|c|}
\hline
Particles&Spin&${\rm{Mass \: Squared}}$\\ \hline
$\mathbf{\Phi^0, \Phi^P, \Phi^+, \Phi^{++}, \Phi^-, \Phi^{--}}$&$0$&$\frac{2{m_H}^2f^2}{v^2(1-(\frac{4v'f}{v^2}))^2}$\\
\hline
$\mathbf{T, T^c}$&$\frac{1}{2}$&$\frac{v^2}{{m_t}^2}(\lambda_1 \lambda_2 f)^2$ \\ \hline
$\mathbf{A_H}$&$1$&${m_z}^2{s_w}^2(\frac{f^2}{5{s'}^2{c'}^2v^2}-1)$\\ \hline
$\mathbf{Z_H}$&$1$&${m_w}^2(\frac{f^2}{s^2c^2v^2}-1)$\\ \hline
$\mathbf{{W_H}^+, {W_H}^-}$&$1$&${m_w}^2(\frac{f^2}{s^2c^2v^2}-1)$\\
\hline
\end{tabular}
\caption{\label{lhparticles} New particles in the Littlest Higgs model~\cite{Han}\cite{Buras}
, where $m_w=\frac{gv}{2}$, $m_z=\frac{gv}{2c_w}$.}
\end{table}

The enlarged electroweak space of the Littlest Higgs model is the 
symmetric tensor representation of $SU(5)$, which is broken at a scale $f$ by a tensor vacuum 
expectation value to the coset
$SU(5)/SO(5)$.  The result is $24-10=14$ Goldstone bosons, which are parameterized by a $5\times5$ non-linear
sigma field $\Sigma(x)$.  Four of these Goldstone bosons will become the complex Higgs doublet of the Standard
Model, four will become the longitudinal modes of four new massive gauge bosons, and the remaining six will form a 
massive complex scalar triplet. This scalar triplet has a positive mass squared, which we require~\cite{Han}, only so long as its
vacuum expectation value $v'$ satisfies (See Table \ref{lhparticles}):

\begin{eqnarray}
v'<\frac{v^2}{4f} \approx \frac{0.015\: {\rm{TeV}}^2}{f}
\end{eqnarray}
where $v=246$ GeV is the Standard Model electroweak scale.  In this work $v'$ enters only in the determination of the lightest
new particle and is set to $2$ GeV, which satisfies
the above relation as long as $f$ is less than $7.5$ TeV. 

The $SU(5)$ group contains an $SU(2)_1 \otimes U(1)_1 \otimes SU(2)_2 \otimes U(1)_2$
subgroup, which is gauged,\footnote{A variant of the Littlest Higgs model omits the gauging of one of the $U(1)$ groups, which is 
acceptable on the grounds that the contribution to quadratic divergence by the photon loop is numerically small.  
This variant is increasingly favored in the literature, since direct search limits
would be greatly eased, along with constraints from electroweak precision data, at the cost of an additional ad hoc 
assumption.  The ungauged $U(1)$ group is then expressed as an additional scalar with interesting phenomenology~\cite{Rain}.}
with gauge couplings $g_1, g_1', g_2,$ and $g'_2$, respectively.  The result is eight
gauge bosons, which after diagonalization to mass eigenstates are the gauge bosons of the Standard 
Model, here labeled $A_L$, $Z_L$, $W_L^{\pm}$, and new gauge bosons with masses of order $f$, labeled $A_H$, $Z_H$,
$W_H^{\pm}$.  The mass diagonalization can be described in terms of the mixing angles:

\begin{eqnarray}
c = {\rm{cos}}\:\theta= \frac{g_1}{\sqrt{{g_1}^2+{g_2}^2}} = \sqrt{1-s^2}
\end{eqnarray}
and
\begin{eqnarray}
c' = {\rm{cos}}\:\theta'= \frac{{g'}_1}{\sqrt{{{g'}_1}^2+{{g'}_2}^2}} = \sqrt{1-{s'}^2}.
\end{eqnarray}
The allowed range of mixing angles is bounded by the perturbative limits

\begin{eqnarray}
g_1,g_2,g'_1,g'_2 < 4\pi.
\end{eqnarray}

With the $SU(5)$ represented in this way, quadratic divergences of Higgs mass due to Standard Model 
gauge boson loops are cancelled by opposite-sign contributions from the new massive gauge bosons, leaving a 
logarithmic leading contribution term. 

\begin{table}[t]
\begin{tabular}{|c|c|c|}
\hline
$f$&$SU(5)/SO(5)$ symmetry-breaking vev& $0.5$ to $6.0$ TeV\\ \hline
$c$&Cosine of the $SU(2)_1 \otimes SU(2)_2$ mixing angle& $0.05$ to $0.99$\\ \hline
$c'$&Cosine of the $U(1)_1 \otimes U(1)_2$ mixing angle& $0.03$ to $0.99$\\ \hline
$\lambda_1$&Top sector Yukawa parameter&$1.0$\\ \hline
$y_u$&quark hypercharge assignment parameter&$-\frac{2}{5}$\\ \hline
$y_e$&lepton hypercharge assignment parameter&$+\frac{3}{5}$\\ \hline
$v'$&scalar triplet vev&$2.0$ GeV\\ \hline
\end{tabular}
\caption{\label{lhparameters}  Free parameters of the Littlest Higgs model~\cite{Han}, with typical values
used in this study.  The parameters $c$ and $c'$ are bounded by perturbative limits.}
\end{table}

To address the largest contribution to the quadratic divergence, that of the top quark loop, a new pair of 
massive top-like fermions $\tilde{t}$, $\tilde{t}^C$ is introduced.  A row vector with 
$\tilde{t}$ is constructed as $\chi = (b_3 \: t_3 \: \tilde{t})$, where ($t_3$ $b_3$) is the 
third-generation quark weak doublet.  The coupling of
$\chi$ to the Higgs boson is mediated by the Yukawa terms in the Lagrangian:

\begin{eqnarray}
\frac{1}{2} \lambda_1 f \epsilon_{ijk} \epsilon_{xy} \chi_i \Sigma_{jx} \Sigma_{ky} {u'_3}^C 
+ \lambda_2 f \tilde{t} \tilde{t}^C + {\rm{higher\, corrections}}
\end{eqnarray}
where $i,j,k$ are summed over $1,2,3$ and $x,y$ are summed over $4,5$, so as to pick the Higgs doublet out
of $\Sigma (x)$.  The second term sets the overall mass scale of the new fermions
to $f$.  After mass diagonalization in this sector, the Standard Model top and bottom quarks emerge along with
new massive top partners, which cancel the leading quadratic divergences.

Because the Standard Model top quark mass is known, the parameters $\lambda_1$ and $\lambda_2$ are constrained.  
The constraint works out to be:
\begin{eqnarray}
\frac{1}{{\lambda_1}^2}+\frac{1}{{\lambda_2}^2} \approx \left( \frac{v}{m_t} \right)^2.
\end{eqnarray}
Unless otherwise indicated, $\lambda_1$ is set to $1$, with $\lambda_2$ set as required to produce the top mass
$m_t=175$ GeV. 

The requirement that the Yukawa terms be gauge-invariant restricts the $U(1)_1$ and $U(1)_2$ ``hypercharge" 
assignments of the fermions, such that the sum for each fermion equals the Standard Model
hypercharge~\cite{Han}.  The remaining freedom of assignment is limited to one parameter for the quarks, $y_u$,
and one for the leptons, $y_e$.  This freedom is eliminated if we require that all anomalies are cancelled in 
the theory, resulting in the fixed values $y_u = -\frac{2}{5}$ and $y_e = \frac{3}{5}$.  For the top quark, these
assignments are fixed by the Yukawa terms above~\cite{Csaki}, but in general anomaly cancellation is not strictly required 
since this theory is explicitly not complete, and additional anomaly-cancelling contributions could 
arise from unidentified higher-energy contributions.  The result is a proliferation of more or less unmotivated
assignment choices.  Motivation for specific choices can be found in the various high-energy completions of the 
theory, in which all anomalies should cancel.  In this work the anomaly-cancelling choices will
be made in all cases for third-generation fermions, and for all fermions unless noted.

The free parameters of the Littlest Higgs model and typical values used in this study are summarized in Table 
\ref{lhparameters}.    

\section{DIRECT SEARCH LIMITS}

\begin{figure}[t]
\epsffile{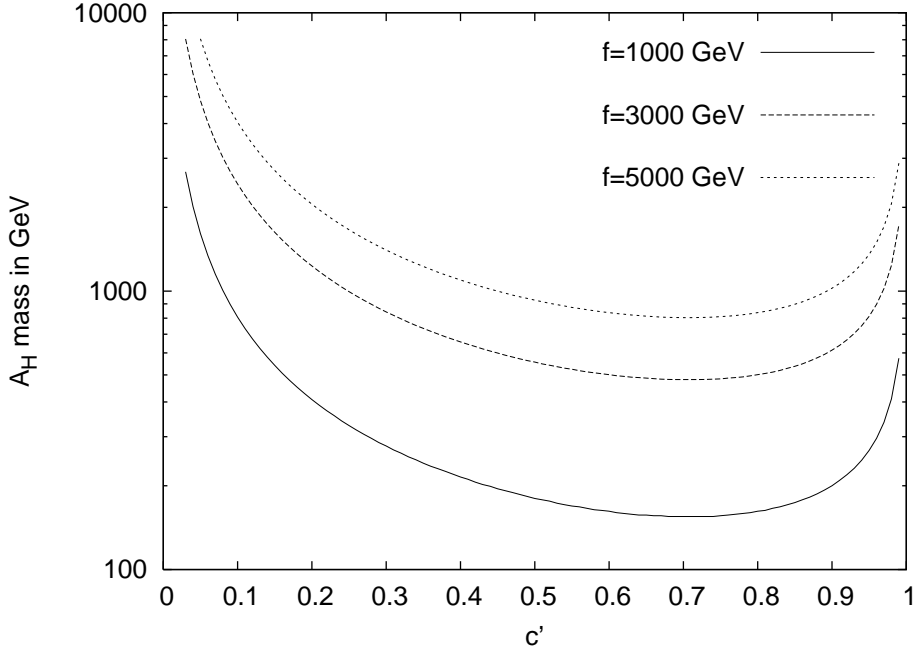}
\caption{\label{ahmass} Mass of the $A_H$, for various scales $f$.}
\end{figure}

To determine the direct-search exclusion limits for the Littlest Higgs model, as a function of the free parameters of
the theory, it is useful to begin by identifying the lightest new particle, since lighter particles are generally 
easier to produce and therefore to observe, provided that they do not have suppressed couplings to Standard Model particles.
As has been widely noted 
in the literature, the $A_H$ is the lightest new particle for most parameter choices.  A detailed scan of the 
$f,c,c'$ parameter space shows that the $A_H$ is always the lightest particle for $c'>0.12$.  For $c'<0.12$, there are
parameter choices for which the $\Phi$ and for which the degenerate $W_H, Z_H$ are the lightest new particles.
The direct search limits on the Littlest Higgs model will be found from 
applicable searches for new massive particles.  Since all the new particles of the Littlest Higgs model have masses proportional to
$f$ in leading order, I will use $f$ as a convenient common measure of the lower exclusion limits on the model.   

The mass of the $A_H$ is displayed in Figure \ref{ahmass} as a function of $c'$ for
several values of $f$.  It scales with $f$, and is strongly
dependent on $c'$.  The current Tevatron run II search reach for narrow resonances
similar to $A_H$ is about $900$ GeV~\cite{cdf05}, so current direct search data will be able to rule out much of parameter space at 
$f=3$ TeV and below, but less parameter space above $f=3$ TeV.  

Since the $A_H$ is in most of parameter space the lightest new particle,
only its decays to Standard Model particles will be 
considered.  For $c' < 0.12$, decays to other new Littlest Higgs model particles, such as 
$A_H \to W_H + W_L$, are possible for some parameter choices and would need to be included in a more
detailed study of this portion of parameter space.  

The $A_H$ can decay to Standard Model fermion pairs with partial width:
\begin{eqnarray}
\Gamma (A_H \to f\bar{f}) = \frac{C\:M_A}{12\pi}
(1-4\frac{{M_f}^2}{{M_A}^2})^{\frac{1}{2}}
({g_v}^2(1+2\frac{{M_f}^2}{{M_A}^2})
+{g_a}^2(1-4\frac{{M_f}^2}{{{M_A}^2}}))
\end{eqnarray}
where $C$ is the appropriate color factor for the fermions, $M_A$ is the $A_H$ mass, $M_f$ is the fermion mass,
and $g_v$ and $g_a$ are the appropriate vector and axial couplings of the $A_H$ to the fermions, as catalogued in~\cite{Han}.

The decay width of a particle affects how and to what extent it is experimentally detectable, since 
production of particles
with very large decay widths may be difficult to distinguish from background processes.
In the case of a supposed new particle,
it is useful to characterize the decay width as a function of the unknown new particle mass.  For the $A_H$,
the decays to fermion pairs are to leading order manifestly linear in $A_H$ mass.

The $A_H$ can also decay to Standard Model $W$ bosons with partial width:

\begin{eqnarray}
\Gamma (A_H \to W^+W^-) =g_{WWA}^2\frac{1}{192\pi}
M_A \left( \frac{M_A}{M_W} \right)^4 \left( 1-4\frac{{M_W}^2}{{M_A}^2} \right)^{\frac{3}{2}}
\left( 1+20\frac{{M_W}^2}{{M_A}^2}
+12\frac{{M_W}^4}{{M_A}^4} \right)
\end{eqnarray}
where $g_{WWA}$ is the triple boson vertex factor and $M_W$ is the mass of the Standard Model $W$. 

Note that the mass of the $A_H$ is proportional to $f$, so linearity in $f$ means linearity in $M_A$. 
The $A_HW^+W^-$ coupling constant is: 

\begin{eqnarray}
g_{AWW}=\frac{5}{2}\frac{ec_w}{{s_w}^2}\frac{v^2}{f^2}s'c'({c'}^2-{s'}^2)
\end{eqnarray}
so decays to $W_L$ pairs are suppressed by four powers of $f$ in the coupling constant squared,
which offset four of the five powers of $f$ in the partial width (from the powers of $M_A$), so the partial
width to $W$s is also linear in $M_A$.
\footnote{In a previous work~\cite{Park}, the possibility of decays to $W$ pairs
was neglected due to an apparent belief that the highly suppressed coupling would make the partial width negligible.
That is in fact not the case, as is shown here:  Kinematic factors (the polarization sums) 
in the amplitude restore the partial width to linearity
in $M_A$.  That work also has a typographically incorrect expression for their kinematic factor $\lambda$.}

\begin{figure}[t]
\epsffile{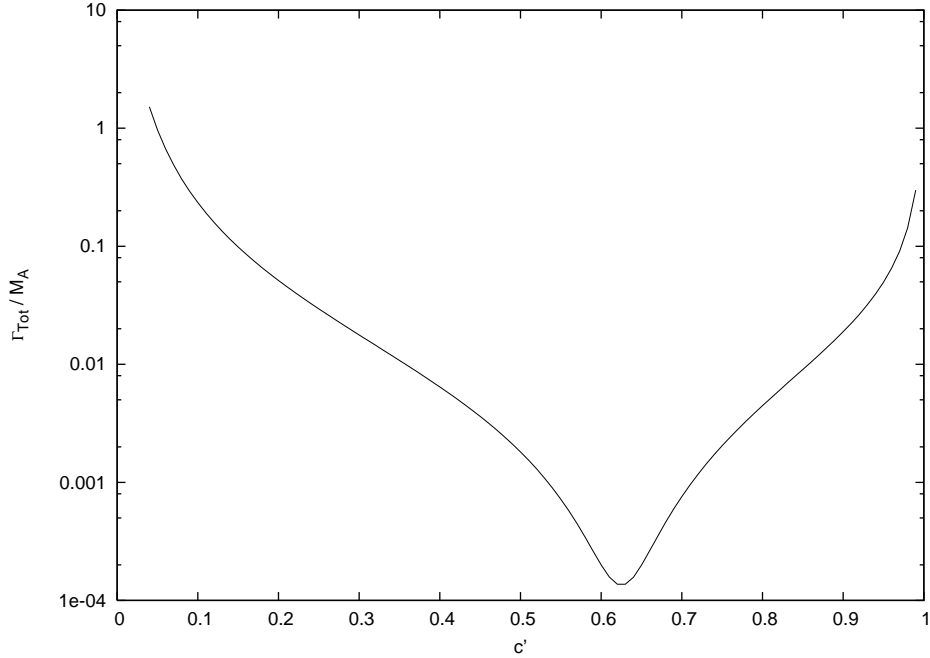}
\caption{\label{widthrat} The ratio of the total decay width of the $A_H$ to its mass in the Littlest Higgs model, 
calculated for $f=3000$ GeV and $c=0.1$.  The ratio is only very weakly dependent on $f$ and $c$, 
as long as the $A_H$ is above the top pair production threshold, because the
decay width is linear in the mass.  Only decays to Standard Model particles are included, so the curve represents a
lower limit for $c' < 0.12$.  Higher-order terms in the Sigma field expansion are not included and may significantly
modify the large width regions.}
\end{figure}

Decays are also possible to a $Z^0$ boson and Higgs boson:

\begin{eqnarray}
\Gamma (A_H \to Z^0H) =\frac{{g'}^2{\rm{cot^2}}(2\theta')|\vec{P}|}{96\pi}
\left( 8 \frac{{M_Z}^2}{{M_A}^2} + \left( 1 + \frac{{M_Z}^2}{{M_A}^2} - 
\frac{{M_H}^2}{{M_A}^2} \right)^2 \right)
\end{eqnarray}
where $|\vec{P}|$ is the momentum of one of the outgoing decay products:

\begin{eqnarray}
|\vec{P}|=\frac{M_A}{2}\sqrt{1+ \frac{{M_Z}^4}{{M_A}^4} + \frac{{M_H}^4}{{M_A}^4} -
2 \frac{{M_Z}^2}{{M_A}^2} - 2\frac{{M_H}^2}{{M_A}^2} - 2\frac{{M_Z}^2{M_H}^2}{{M_A}^4}}.
\end{eqnarray}

In the large $M_A$ limit ($|\vec{P}| \to \frac{M_A}{2}$),
decays to $Z^0H$ are also linear in $M_A$. The dominant decays are therefore all linear in $M_A$ and so the
total width is as well. Note that in the limit of massless decay products, the partial widths 
to $W^+W^-$ pairs and to $Z^0H$ are equal, as they must be in accordance with the 
Goldstone boson equivalence theorem. However, at sufficiently low $A_H$ masses,
where the decay product masses are not negligible, the partial widths can be quite different.  

\begin{figure}[t]
\epsffile[72 54 450 306]{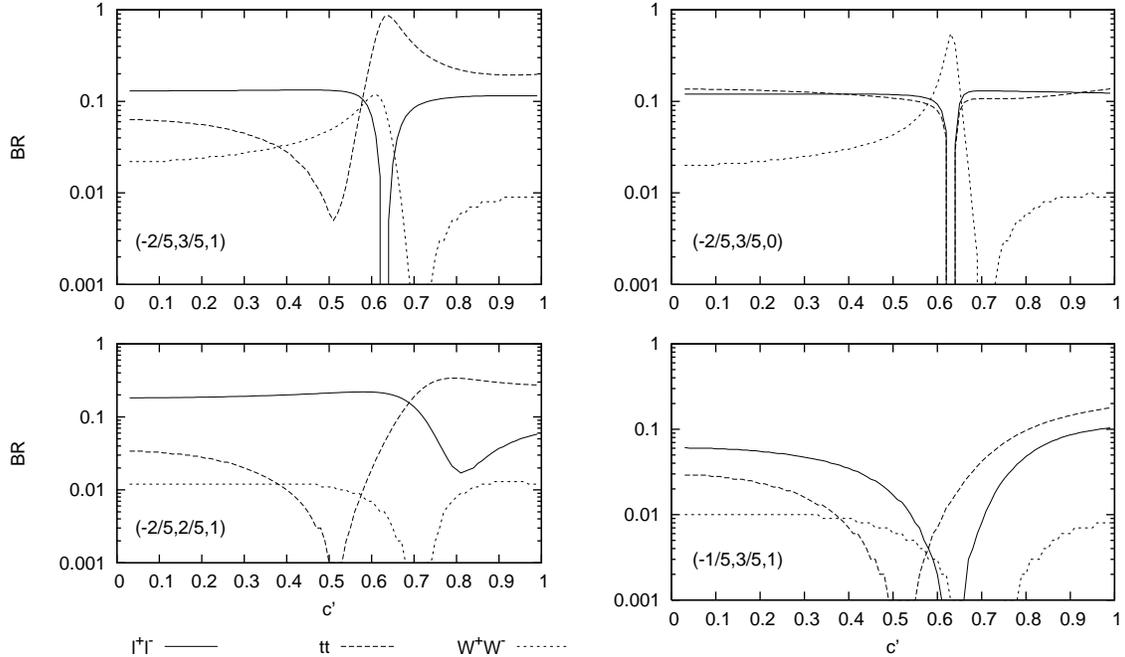}
\caption{\label{ahbranch} Some branching ratios for the decays of the Littlest Higgs $A_H$, for various choices of 
the free parameters ($y_u$,$y_e$,$\lambda_1$). The $l^+l^-$ branching ratio is for each of $e^+e^-,\mu^+ \mu^-, \tau^+ \tau^-$.
The choice of $\lambda_1=0$ will not correctly recreate the top quark
mass; this choice was included to allow comparison to a previous study.}
\end{figure}

There would also be, for sufficient $A_H$ mass, the quartic decays $A_H \to Z^0HH$,
$A_H \to \gamma W^+ W^-$, and $A_H \to Z^0 W^+ W^-$.  These decays are all suppressed by a 
four powers of the coupling constants, and so would likely be rare.  They are not considered further here. 

The resulting total width-to-mass ratio of the $A_H$ is presented in Figure \ref{widthrat}. Depending on the value of the mixing
angle $c'$, the $A_H$ can range from very wide, with a total width in excess of the mass, to extraordinarily narrow.  For example,
at $c'=0.1$, the total width is $568$ GeV at a mass of $2425$ GeV, but for $c'=0.5$, the total width is only $1.0$ GeV at a mass
of $556$ GeV.  In general, the very large
widths occur for large ($c'\gtrsim 0.9$) and small ($c' \lesssim 0.15$) cosines of the $\theta '$ mixing angle where,
 as we have seen, the $A_H$ mass becomes large. It may be that higher-order terms in the Sigma field would significantly
modify the width in the large-width regions; this possibility is not explored here. 
For very large widths, the Breit-Wigner propagator would need a fully momentum-dependent width, but 
that is not done here since the large $A_H$ masses where this would occur are beyond current search limits in any case.
The region of very narrow width ($0.5 \lesssim c' \lesssim 0.7$) occurs because
in this region, the coupling to fermions, and so the largest contributions to the total width, approach zero, as will
be considered in the discussion of branching ratios below.  The essential point is that for almost all of the range of 
$c'$, $\Gamma_{Tot} << M_A$, so we can find exclusion limits from generic searches for narrow resonances.

The branching ratios of the $A_H$ decay are shown in Figure \ref{ahbranch}, for four sets of choices of the hypercharge
and Yukawa parameters.  For the $y_u=-\frac{2}{5}$, $y_e=\frac{3}{5}$ hypercharge assignments, which cancel all Standard
Model anomalies, the decays to dileptons $A_H \to e^+e^-$ and $A_H \to \mu^+\mu^-$ are dominant for more than half 
of the range of $c'$.   The dilepton decays are in fact large for all values of $c'$ and all four studied combinations
of the parameters $y_u,y_e,\lambda_1$ except, in three cases, for a narrow region around $c'=0.63$.  This ``leptophobic"
range of parameters is due to a special cancellation of terms in the vector and axial couplings of the $A_H$ to the leptons~\cite{Han}:

\begin{eqnarray}
g_v(A_Hl\bar{l})&=&\frac{g'}{2s'c'}(2y_e-\frac{9}{5}+\frac{3}{2}{c'}^2),\nonumber\\
g_a(A_Hl\bar{l})&=&\frac{g'}{2s'c'}(-\frac{1}{5}+\frac{1}{2}{c'}^2).
\end{eqnarray}
Inspection of the couplings to all Standard Model fermions (other than top) shows that they, too, vanish at this special 
angle, as they must in order to maintain anomaly cancellation. To constrain this region we can consider decays to 
top pairs, or to $W$ bosons. 
Couplings to top pairs do not vanish at this same angle
because of the presence of an additional coupling term originating in the mixing of the top quark with the new fermion
$\tilde{t}$:
\begin{eqnarray}
g_v(A_Ht\bar{t})&=&\frac{g'}{2s'c'}(2y_u+\frac{17}{15}-\frac{5}{6}{c'}^2
-\frac{1}{5}\frac{\lambda_1^2}{\lambda_1^2+\lambda_2^2}),\nonumber\\
g_a(A_Ht\bar{t})&=&\frac{g'}{2s'c'}(\frac{1}{5}-\frac{1}{2}{c'}^2
-\frac{1}{5}\frac{\lambda_1^2}{\lambda_1^2+\lambda_2^2}).
\end{eqnarray}

\begin{figure}
\epsffile{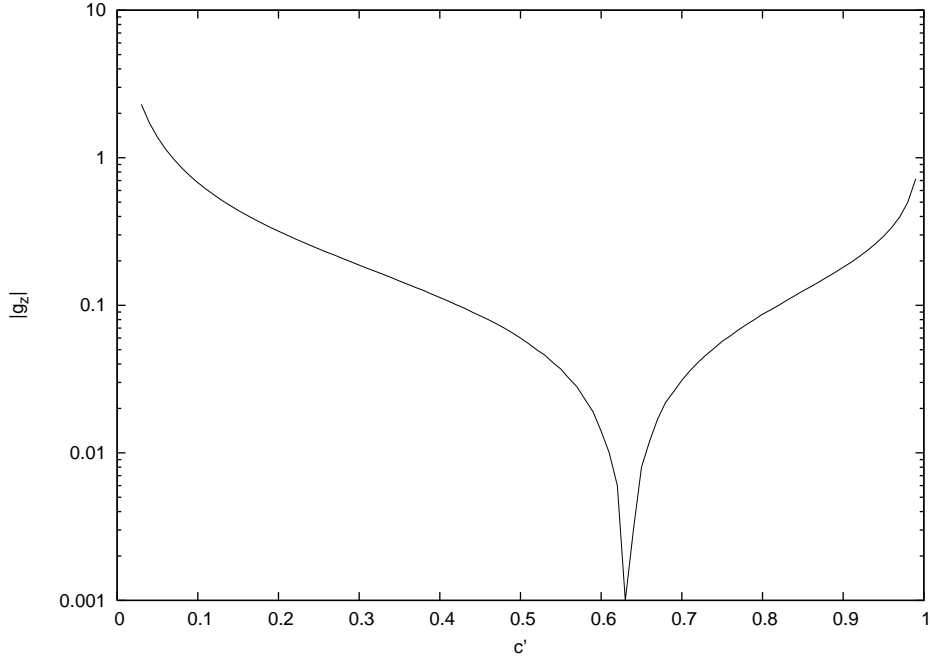}
\caption{\label{gz} The effective chiral coupling constant $|g_z|$ in the Littlest Higgs model.}
\end{figure}

The most recent dilepton resonance results are from CDF~\cite{cdf06}, in which non-model specific searches
using $450\: {\rm{pb}}^{-1}$ of integrated luminosity from Run II of the Tevatron are presented 
in terms of the generic $U(1)$ gauge boson families of Carena, Daleo, Dobrescu, and Tait (CDDT)~\cite{cddt}. This latest
experimental result, as it is presented, is somewhat less useful in constraining the Littlest Higgs model
than prior studies, which presented generic cross-section limits.  One reason for this is that the CDDT framework defines its gauge boson 
families so as to achieve full anomaly 
cancellation,
which does not occur for effective theories with incomplete anomaly cancellation, as can be the case here.  For the special
case of complete anomaly cancellation, we can map the vector and axial couplings of the $A_H$ to the chiral couplings of 
CDDT by subtracting and adding, and find that the $A_H$ is a member of the CDDT family $q+xu$, with $x=4$. For the couplings
to $u\bar{u}$, for example, we find the left-handed coupling to be:

\begin{eqnarray}
g_zq_L=g_v-g_a=\frac{g'}{2s'c'}(2y_u+\frac{14}{15}-\frac{1}{3}{c'}^2)=\frac{g'}{2s'c'}(\frac{2}{15}-\frac{1}{3}{c'}^2)
\end{eqnarray} 
which, for $q+4u$, is
\begin{eqnarray}
      &=&\frac{1}{3}g_z
\end{eqnarray}
so
\begin{eqnarray}
g_z&=&\frac{3g'}{2s'c'}(\frac{2}{15}-\frac{1}{3}{c'}^2).
\end{eqnarray}

Likewise, for the right-handed coupling:
\begin{eqnarray}
g_zu_R=g_v+g_a=\frac{g'}{2s'c'}(2y_u+\frac{4}{3}-\frac{4}{3}{c'}^2)=\frac{4}{3}g_z.
\end{eqnarray}
The remaining $A_H$ couplings are consistent with this identification.  Note that the identification with any CDDT family
will fail unless we make the anomaly-cancelling choices $y_u=-\frac{2}{5}$, $y_e=\frac{3}{5}$,
and that the other new neutral gauge boson in the
Littlest Higgs model, the $Z_H$, does not conform to any CDDT family, because it is not a $U(1)$ gauge boson.  In these cases constraints 
could be found from the limits on generic $c_u$ and $c_d$ couplings, as also defined by CDDT~\cite{cddt}.

\begin{figure}
\centering
\epsfxsize=3.0in
\epsffile[176 50 410 302]{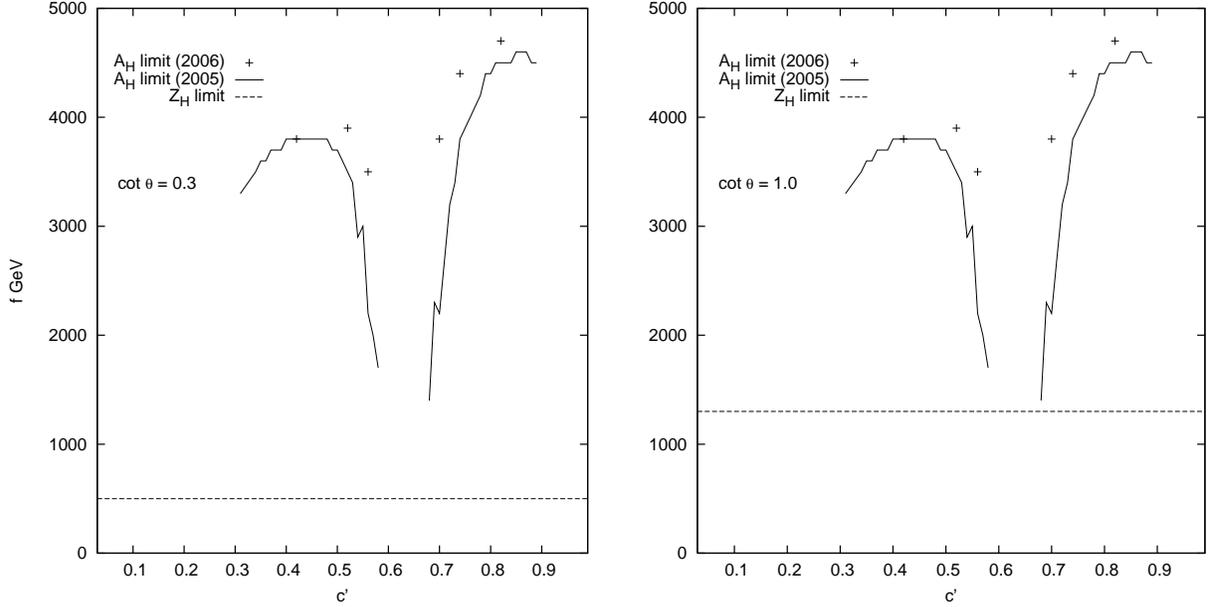}
\caption{\label{lhlimits} 
The direct search limits on $f$ for the Littlest Higgs Model, with ${\rm{cot}}\:\theta = 0.3$ on the left,
and ${\rm{cot}}\:\theta = 1.0$ on the right, for the anomaly-cancelling choice of free parameters
$y_u = -\frac{2}{5}, y_e = \frac{3}{5}$.  Regions below the points and curves are excluded.  The six points
are limits from a search~\cite{cdf06} specific to the CDDT framework\cite{cddt}.  The $Z_H$ curves are developed
from search data also presented in~\cite{cdf06}.  The $A_H$ curves are developed from data presented in~\cite{cdf05}.}
\end{figure}

Numerical values of $|g_z|$ are presented in Figure \ref{gz}.  The absolute value is shown since there is an inconsequential
sign change in the expression for $g_z$ at $c'=0.63$.  CDF presents results for
three discrete values of the coupling:  $g_z = 0.03, 0.05, {\rm{and}}\: 0.10$.  It is therefore possible to constrain the Littlest Higgs
model at a discrete set of six values of $c'$.  Those limits are presented in Figure \ref{lhlimits}, as a set of six values of $f$, one
for each available value of $c'$.  These limits range as high as $f=4700$ GeV.  From these six points we find a 
suggestion of the limits for the full range of $c'$.  A presentation by CDF of limits at a larger and denser set of $g_z$ values would 
permit a more detailed limit on $f$, for the anomaly-cancelling case.  

A fuller picture can be formed by using generic search results for resonances in dileptons, as 
done in a previous study using Tevatron Run I data and a fixed excluded event rate for all $A_H$ masses~\cite{Hewett}.
The most recent
data set is from CDF~\cite{cdf05} for $200\: {\rm{pb}}^{-1}$ of integrated luminosity from Run II of the Tevatron.  Generic 
resonance searches for particles of spin $0\:$,$1$, and $2$, with masses from $150$ to $900$ GeV, are presented as $95\%$ 
confidence limits on $\sigma(p\bar{p} \to X) \cdot {\rm{BR}}(X \to l\bar{l})$.  Observed limits for spin $1$ particles
range from $21$ fb to $490$ fb. 

\begin{figure}[t]
\epsffile[72 54 450 306]{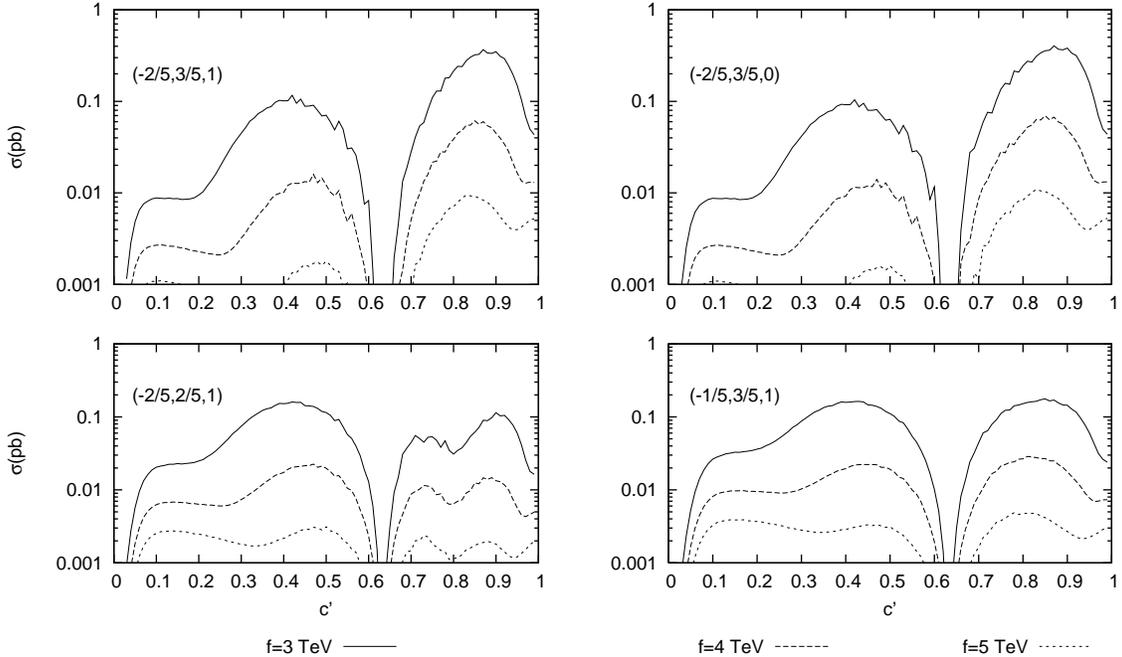}
\caption{\label{lhllsigma} The cross section $\sigma (p\bar{p} \to A_H \to l\bar{l})$ in the Littlest Higgs model,
for several choices of the parameters ($y_u$,$y_e$,$\lambda_1$), and symmetry-breaking scale $f=3,4,$ and $5$ TeV.
The small jags in some curves are artifacts of the Monte Carlo calculation.
}
\end{figure}

General limits on the Littlest Higgs model can be found by calculating the cross section 
$\sigma(p\bar{p} \to A_H \to l\bar{l})$ and comparing to the search limits on narrow spin-$1$ resonances for various
values of $f$, and so various $M_A$.  The cross-sections themselves for a few values of $f$ and other free parameters
are presented in Figure \ref{lhllsigma}.  These cross-sections are calculated at Tevatron Run II energy $\sqrt{s}=1.96$ TeV,
at leading order with a K-factor correction:
\begin{eqnarray}
K=1+\frac{4}{3}\frac{\alpha_s (Q^2)}{2\pi}(1+\frac{4}{3}\pi^2)
\end{eqnarray}
applied as a function of $Q^2$ over the resonance,
and using the CTEQ6L parton distribution functions~\cite{cteq}.  The cross-section is relatively large
around $c'=0.4$, and $c'=0.9$, so we will find the most stringent limits near these values.  The cross-section
becomes vanishingly small for large and small $c'$ because the $A_H$ mass becomes too large for significant production
at Tevatron energies. Around $c'=0.63$, the cross-section vanishes due to decoupling from leptons and quarks.  

The results of comparing the computed cross-sections to the CDF search limits are presented in Figure \ref{lhlimits}.  
The overall symmetry-breaking scale $f$ is found to be excluded up to as much as $4.5$ TeV, in the anomaly-cancelling case,
for some values of $c'$.  For other values of $c'$,
the model is not constrained, due to either decoupling or to a large $A_H$ mass.  The calculated exclusion curves are somewhat
below the exclusion points provided by CDF's CDDT-specific analysis, as would be expected since the calculated curves are 
based on less than half of the 
integrated luminosity used to generate the CDDT-specific points.  This relationship 
provides a cross-check on the independent Monte Carlo
simulations of the $A_H$ cross section described here with that done by CDF.  It is not possible to seek to constrain
the Littlest Higgs model much further in the high mass regions of parameter space with Tevatron data. 

\begin{figure}
\epsffile[72 54 450 306]{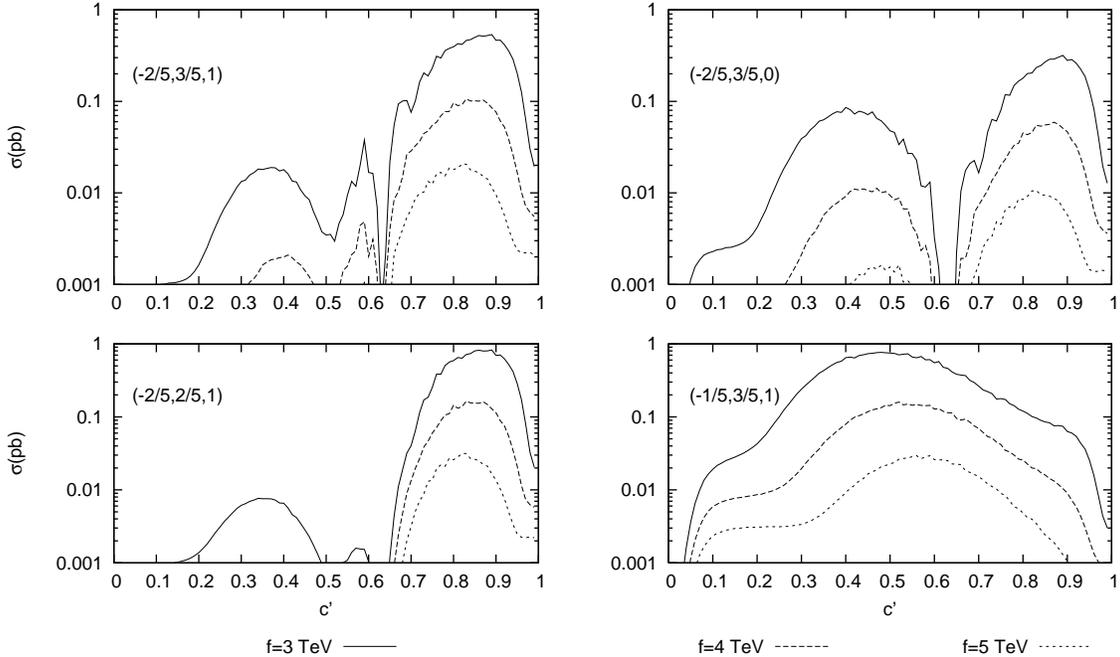}
\caption{\label{lhttsigma} The cross section $\sigma (p\bar{p} \to A_H \to t\bar{t})$ in the Littlest Higgs model,
for several choices of the parameters ($y_u$,$y_e$,$\lambda_1$).
The small jags in some curves are artifacts of the Monte Carlo calculation.}
\end{figure}  

To attempt to constrain the Littlest Higgs model in the decoupling range around $c'=0.63$, we can look to 
the current limits on resonances in top pair production by \D0 and CDF~\cite{tt}.  The \D0 results, for $680
\:{\rm{pb}}^{-1}$ of integrated luminosity of Run II data, are the most useful for this purpose since generic narrow
resonance search limits are presented.  The \D0 study assumes a neutral boson width $\Gamma = 0.012 M$,
with $95\%$ confidence limit exclusions ranging from $5.08$ pb at a mass of $350$ GeV to a minimum of 
$0.88$ pb at $650$ GeV, and then increasing to $1.43$ pb at $1000$ GeV.  These limits are considerably larger
than the equivalent results for leptons, so we cannot expect as strong a constraint overall. 
The comparable cross section $\sigma(p\bar{p} \to A_H \to t\bar{t})$ is calculated as previously for the 
cross-section to leptons, and is presented in Figure \ref{lhttsigma}.  For the anomaly-cancelling case, the 
calculated cross-section is at most $0.54$ pb, so unfortunately the current experimental limits on resonances
in top pair production do not quite provide a constraint.  For other choices of the free parameters the top
pair resonance limits do provide a constraint for a limited range of $c'$. 

The remaining hope for constraining the decoupling range from $A_H$ decays falls to the ${W_L}^+{W_L}^-$ decay mode.  
No experimental results on resonances in $W$ pair production are available, so we must turn to the total ${W_L}^+ {W_L}^-$ production
cross section.  The most current result in provided by CDF~\cite{ww}, with an observed cross-section of 
$13.6\: \pm 2.3{\rm{(stat)}}\: \pm 1.6{\rm{(sys)}} \pm 1.2{\rm{(lum)}}$ pb.  Since the entire cross-section of $A_H$
production at the Tevatron is at most $4\:$pb~\cite{Han}, the current limits on $W$ pair production are in fact not
able to provide a constraint for the Littlest Higgs model.  

Having exhausted the likely signals for the $A_H$, we turn to one of the next-lightest particles, the $Z_H$.  CDF
has presented a set of model-specific search limits for the Littlest Higgs $Z_H$ for a set of values of 
${\rm{cot}}\:\theta$~\cite{cdf06}.  These limits can be converted to limits on the scale $f$, and 
are displayed in Figure \ref{lhlimits}.  
The $Z_H$ results provide an exclusion limit for all values of $c'$, filling in the $A_H$ decoupling and high mass
regions, although an a much lower exclusion limit.\footnote{To the extent that an incorrect expression for the $Z_H$ mass
has been used in these CDF limits, they may require reconsideration.  My own studies of the effect of correcting the $Z_H$
mass expression suggest that any changes of limits due to this specific correction are likely to be small.}     

\section{CONCLUSIONS}
In this work I have made a fairly thorough study of the direct limits on the Littlest Higgs model provided by current 
searches for new particles, for a range of free parameters, although primarily focused on the 
anomaly-cancelling case.  Recent experimental results permit some constraint in all cases.  For a limited 
range of parameters, lower limits on the symmetry-breaking scale can range as high as $4.7$ TeV.  

The possibility of detection of new particles, and the best means to do so, depends critically on the width of the associated
resonance.  Wide resonances can be difficult to detect, and so determining the width behavior is important.  The survey of
dominant Standard Model decays of the $A_H$ presented here, with all final masses included, has found that all dominant contributions
have partial widths proportional to the $A_H$ mass. 
Large widths do occur for large ($c' \gtrsim 0.9$) and small ($c' \lesssim 0.15$) cosines of the 
$U(1)_1 \otimes U(1)_2$ mixing angle $c'$, and in these regions contributions from non-Standard Model decays, and higher-order
contributions from the $\Sigma (x)$ field may become important. 

As can be seen in Figures \ref{lhlimits} and \ref{lhllsigma},
 a low-mass $A_H$ neutral gauge boson can be present and yet undetectable
in dilepton studies for a significant range of parameter space, due to decoupling from Standard Model leptons.  
The $A_H$ provides a striking counterexample to 
broad claims
that a new neutral gauge boson (sometimes generically referred to as a $Z'$) is ruled out to a high mass scale.  

Some observations can be made on recent presentations of experimental results.  Searches
for individual model-specific new particles are very useful where applicable.
However, in a model such as the Littlest Higgs, with a number of free parameters, carefully described
searches for generic narrow resonances, as is done in~\cite{cdf05}, are also useful.  
Classifications of new particles such as the CDDT framework can play a useful role, too, although this study
provides examples of the limitations of such a system, since it is not meant to comprehensively covers the full range of possible
new neutral gauge bosons.  In the Littlest Higgs model, the $A_H$ boson in the non-anomaly cancelling case, and the $Z_H$ boson, 
cannot be placed in the CDDT framework.  Limits can be found from constraints on the generic couplings $c_u$ and $c_d$ in all 
these cases, however.  

In this study most of the meaningful constraints were obtained from dilepton resonance studies.  As the Tevatron integrated
luminosity builds over the next few years, we can look forward to dilepton results providing increasingly strong constraints or,
in a more exciting possibility, beginning to characterize an actual high mass resonance.

Top quark resonances do not yet provide a constraint in most cases, but, for the anomaly-cancelling case,
 the current search limits are less than two times greater
than the maximum calculated cross-section.  It appears that at higher integrated luminosity, the search
for resonances in top pair production at the Tevatron may well be able to provide a constraint in this case, so further studies in this 
area would be helpful.  If feasible, searches for resonances in $W$ pair production would also be quite useful. 

\section*{Acknowledgments}

Thanks to Dave Rainwater for orienting me to the Little Higgs literature, and for many helpful conversations. 
Lynne Orr provided a number of useful suggestions. 
Kevin McFarland helpfully discussed with me some of the latest CDF results.  
Bob McElrath provided clarifications to~\cite{Han}, and Puneet Batra kindly provided insights
into the formal foundations of the model.  Andrzej Buras promptly provided details of some results in~\cite{Buras},
which is much appreciated.  Tim Tait provided helpful insights into the CDDT framework. Thanks to an
anonymous reviewer for extensive and thoughtful comments. 
This material is based upon work supported by the Department of Energy under Award Number
DE-FG02-91ER40685.

\end{document}